\documentclass[3p,times]{elsarticle}

\usepackage{ecrc}

\usepackage{epstopdf}

\volume{00}

\firstpage{1}

\journalname{Nuclear Physics A}

\runauth{S. Floerchinger et al.}

\jid{nupha}

\jnltitlelogo{Nuclear Physics A}

\usepackage{graphicx}
\usepackage{subcaption}

\usepackage{amsmath,amssymb}

\begin{document}

\begin{frontmatter}



\title{A perturbative approach to the hydrodynamics of heavy ion collisions}

\author[(a)]{Stefan Floerchinger}
\author[(a)]{Urs Achim Wiedemann}
\author[(b)]{Andrea Beraudo}
\author[(c),(d),(e)]{Luca Del Zanna}
\author[(c),(d)]{Gabriele Inghirami}
\author[(f),(g)]{Valentina Rolando}
\address[(a)]{Physics Department, Theory Unit, CERN, CH-1211 Gen\`eve 23, Switzerland}
\address[(b)]{INFN - Sezione di Torino Via Pietro Giuria 1, 10125 Torino, Italy}
\address[(c)]{Dipartimento di Fisica e Astronomia, Universit\`a di Firenze, Via G. Sansone 1, I-50019 Sesto F.no (Firenze), Italy}
\address[(d)]{INFN - Sezione di Firenze, Via G. Sansone 1, I-50019 Sesto F.no (Firenze), Italy}
\address[(e)]{INAF - Osservatorio Astrofisico di Arcetri, L.go E. Fermi 5, I-50125 Firenze, Italy}
\address[(f)]{INFN - Sezione di Ferrara, Via Saragat 1, I-44100 Ferrara, Italy}
\address[(g)]{Dipartimento di Fisica e Scienze della Terra, Universit\`a di Ferrara, Via Saragat 1, I-44100 Ferrara, Italy}




\begin{abstract}
Initial fluctuations in hydrodynamic fields such as energy density or flow velocity give access to understanding initial state and equilibration physics as well as thermodynamic and transport properties. We provide evidence that the fluid dynamic propagation of fluctuations of realistic size can be based on a background-fluctuation splitting and a systematic perturbative expansion in the fluctuating fields. Initial conditions are characterized by a Bessel-Fourier expansion for single events, event-by-event correlations and probability distributions. The evolution equations can be solved order-by-order in the expansion which allows to study the fluid dynamical propagation of single modes, the study of interaction effects between modes, the determination of the associated particle spectra and the generalization of the whole program to event-by-event correlations and distributions.
\end{abstract}


\end{frontmatter}



\section{Introduction}
\label{intro}
Event-by-event fluctuations in the fluid dynamic fields like energy density $\epsilon$, fluid velocity $u^\mu$, shear stress $\pi^{\mu\nu}$, bulk viscous pressure $\pi_\text{bulk}$ and more general also baryon number density $n_B$, electromagnetic charge density, electromagnetic fields and so on, could provide an interesting opportunity to learn much about the physics of heavy ion collisions. They are governed by universal evolution equations and their study can help to constrain thermodynamic and transport properties of the quark gluon plasma. Moreover, the initial conditions for such kinds of perturbations contain interesting information from early times, i.e. from the initial state directly after the collision and the non-equilibrium dynamics that drives (approximate) local thermalization. 

A theoretical program that aims at understanding these kinds of perturbations has to include different steps, ranging from the characterization of initial conditions at the time when fluid dynamics becomes approximately valid via the propagation of these perturbations though the fluid dynamic regime to the determination of their influence on observables such as particle spectra and harmonic flow coefficients at freeze-out. One strategy to implement this is based on numerical simulations of all these steps. However, our goal here is to develop a more analytic approach from which we expect additional insights, which is less demanding from a numerical point of view and which can supplement the numerical simulations. We will argue that a perturbative treatment that is based on an expansion of the fluid fields around a smooth and symmetric "average" solution of the evolution equations, provides a convenient scheme for this purpose \cite{Floerchinger:2013rya,Floerchinger:2013tya}.

We characterize the initial conditions for the transverse density, e.g. enthalpy density $w$, at the initial time $\tau_0$ by a Bessel-Fourier expansion (see \cite{Floerchinger:2013vua} for a detailed discussion)
\begin{equation}
w(r,\phi) = w_\text{BG}(r) + w_\text{BG}(r) \sum_{m,l} w^{(m)}_l \,e^{im\phi} \;J_m\left(z^{(m)}_l \rho(r)\right).
\label{eq1}
\end{equation}
The function $w_\text{BG}(r)$ is the background enthalpy density distribution corresponding to an conveniently taken, azimuthally symmetric event average. The dimension-less coefficients $w^{(m)}_l$ (sometimes also written as $\tilde w^{(m)}_l$ below) posses an azimuthal wavenumber $m$ and a radial wavenumber $l$. The argument of the Bessel functions $J_m$ contain their $l$'th zero crossing $z^{(m)}_l$ and the positive and monotonously increasing function $\rho(r)$ that maps the relevant interval of radii (either $(0,\infty)$ or $(0,R)$ with some large enough radius $R$) to the the interval $(0,1)$. The coefficients $w^{(m)}_l$ can be related to the more standard eccentricities $\epsilon_m$ in a straight-forward way \cite{Floerchinger:2013vua}. A characterization similar to eq.\ \eqref{eq1} can be also be done for vector and tensor-type objects such as the fluid velocity and the shear stress. 

Ensembles of events can be characterized by a probability distribution for the coefficients $w^{(m)}_l$ or, equivalently, by the set of moments
\begin{equation}
\left\langle w^{(m_1)}_{l_1} w^{(m_2)}_{l_2}  \ldots w^{(m_n)}_{l_n} \right\rangle.
\end{equation}
Arguably, apart from the background field $w_\text{BG}(r)$, these moments (as well as generalizations including other fluid fields and rapidity dependence) contain all information from initial state and early equilibration physics that goes into a fluid dynamic description. On the one side it would be nice to constrain these objects phenomenologically in order to constrain models of the initial state. On the other side, if these correlation functions were known -- or at least some of their features -- one could use this information to constrain the medium properties. This motivated to investigate whether there are some universal properties of such correlation functions that are shared by many or all models of initial state physics. 

An idea in this direction is to consider a model of independent (point) sources \cite{Alver:2006wh,Holopainen:2010gz,Bhalerao:2011bp} where the enthalpy density can be written as a sum of localized contributions,
\begin{equation}
w(\vec x) =  \left[ \frac{1}{\tau_0} \frac{dW_\text{BG}}{d\eta}\right]
 \frac{1}{N} \sum_{j=1}^N \delta^{(2)}(\vec x - \vec x_j).
\label{eq3}
\end{equation}
The random positions $\vec x_j$ are here independently and identically distributed according to some distribution $p(\vec x_j)$ that reflects the collision geometry. For this model it is possible to determine correlation functions or moments of the coefficients $w^{(m)}_l$ analytically, both for central and non-central collisions \cite{Floerchinger:2014fta}. 

It is clear that point-like sources are not realistic. The enthalpy density in eq.\ \eqref{eq3} contains infinitely strong gradients and fluid dynamics (resting upon a derivative expansion) cannot be used to propagate it in time. On the other side one may expect that the statistical properties of long-wavelength modes, i.\ e.\ the moments of coefficients $w^{(m)}_l$ with small $m$ and $l$, are independent of the detailed shape of the sources so that the point-like case is just a convenient limit that allows an analytic treatment. A quite general result, that is actually independent of the shape of the sources, is that connected correlation functions or cumulants of $n$ weights $w^{(m)}_l$ scale with the number of sources $N$ like $1/N^{n-1}$ (see also refs.\ \cite{Yan:2013laa, Bzdak:2013raa} for similar results on eccentricities). For non-central collisions that scaling is broken by terms with known impact parameter dependence \cite{Floerchinger:2014fta}.

Let us now come to the fluid dynamic response. As mentioned above, the idea is here to make a perturbative expansion in deviations from a smooth and symmetric background. If only the enthalpy density fluctuates, this amounts to an expansion in powers of the $w^{(m)}_l$'s. If one combines all fluid dynamic fields into a single field $h=(w,u^\mu,\pi^{\mu\nu}, \ldots)$, one can write them at the initial time $\tau_0$ formally as
$h= h_0 + \epsilon h_1$,
where $h_0$ is the background and $\epsilon h_1$ is the perturbation. If the perturbation is small enough one can expand the fields at later times like $h=h_0 + \epsilon h_1 + \epsilon^2 h_2 + \ldots$ where $h_1$ is the solution of the fluid dynamic equations linearized around the background solution $h_0$ and non-linear corrections like $h_2$, $h_3$ etc.\ can be calculated iteratively. This kind of expansion scheme can also be used at kinetic freeze-out see \cite{Floerchinger:2013hza} for an explicit discussion of the leading, linear order.

It is now relatively easy to see that the azimuthal rotation symmetry of the background implies for the harmonic flow coefficients in response to initial density perturbations the following expansion
\begin{equation}
V_{m}^*  =  v_m e^{-i\, m\, \psi_m} 
= \sum_{l} S_{(m) l} \, w^{(m)}_{l} + \sum_{\substack{m_1,m_2,\\l_1,l_2}} S_{(m_1,m_2) l_1,l_2} \, w^{(m_1)}_{l_1}\, w^{(m_2)}_{l_2}  \, \delta_{m,m_1+m_2}+ \ldots.
\end{equation} 
The objects $S_{(m)l}$ are here the linear dynamic response functions, $S_{(m_1,m_2)l_1,l_2}$ the quadratic dynamic response functions etc. These depend on the thermodynamic and transport properties, in particular the velocity of sound and viscosity. 

Moments of flow coefficients can be written as
\begin{equation}
\begin{split}
\left\langle V^*_{m_1} \cdots V^*_{m_n} \right\rangle = & S_{(m_1)l_1} \cdots S_{(m_n)l_n} \left\langle w^{(m_1)}_{l_1} \cdots w^{(m_n)}_{l_n} \right\rangle  \;\;+\;\; \text{non-linear terms}.
\end{split}
\label{eq4}
\end{equation}
In other words they are determined by a combination of dynamical response functions and correlation functions of initial density perturbations. Interestingly, it follows from a theorem proven in \cite{JamesMayne} that the above mentioned scaling with the number of sources of the density correlation functions for central collisions implies that cumulants of flow coefficients scale in the same way, i.\ e.\ $v_m\{n\}^n \sim 1/N^{n-1}$, even at the full non-linear level of eq.\ \eqref{eq4} \cite{Floerchinger:2014fta}.

\begin{figure}
\begin{center}
\begin{subfigure}[b]{0.46\textwidth}
\includegraphics*[width=\textwidth]{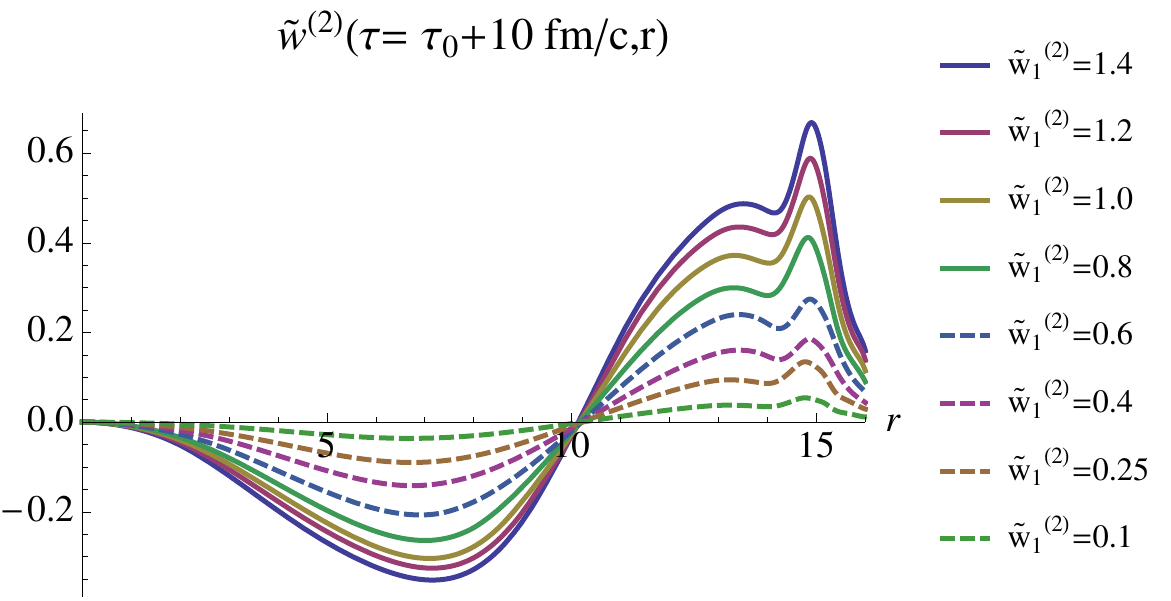}
\caption{}
\end{subfigure}
\begin{subfigure}[b]{0.46\textwidth}
\includegraphics*[width=\textwidth]{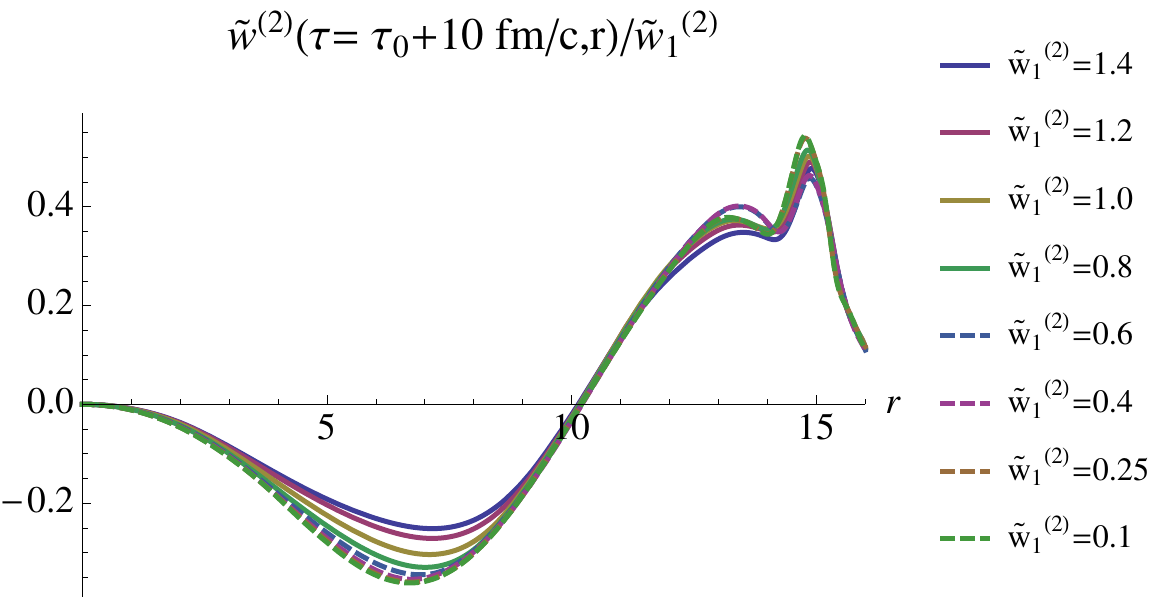}
\caption{}
\end{subfigure} \\[0.3cm]
\begin{subfigure}[b]{0.46\textwidth}
\includegraphics*[width=\textwidth]{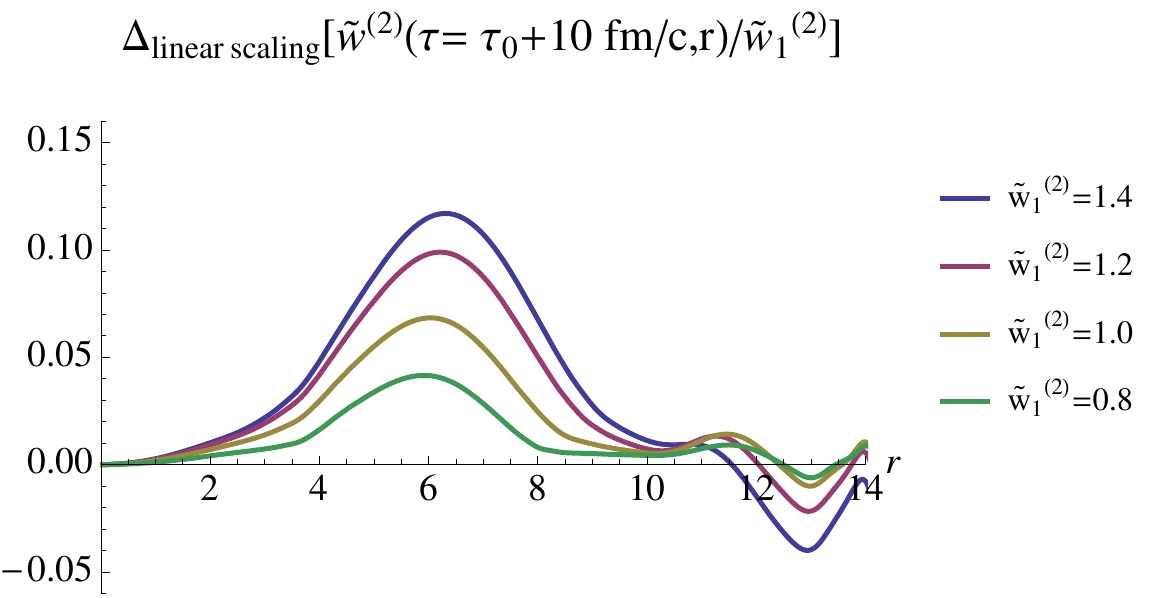}
\caption{}
\end{subfigure}
\begin{subfigure}[b]{0.46\textwidth}
\includegraphics*[width=\textwidth]{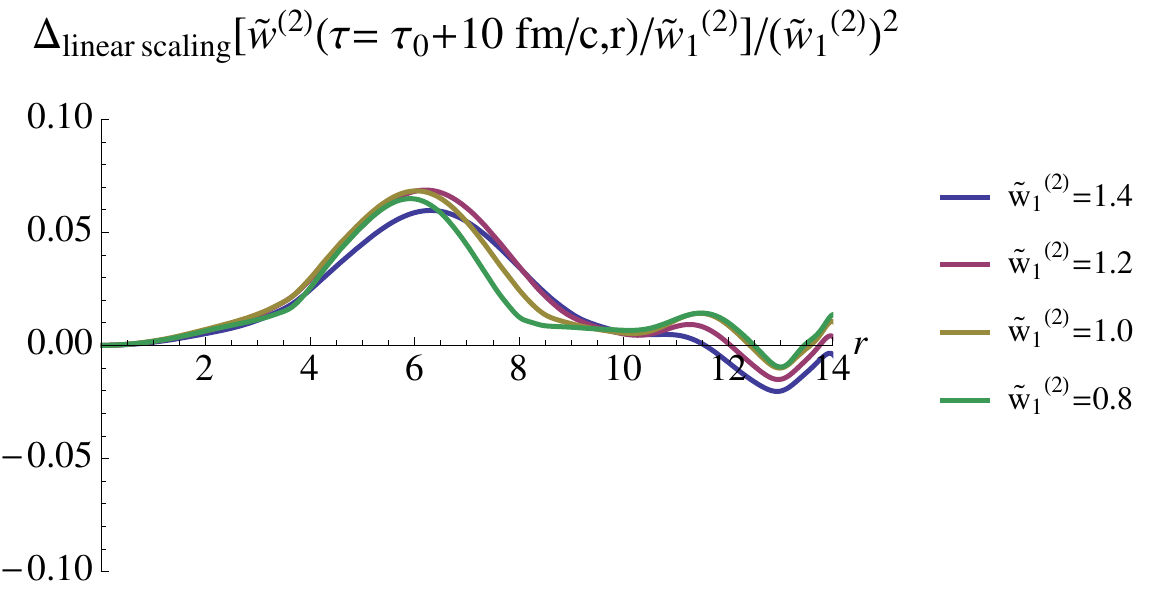}
\caption{}
\end{subfigure}
\caption{Fourier coefficients of the enthalpy density as a function of radius according to eq. \eqref{eq7} at time $\tau= \tau_0+10\, \text{fm/c}$. In (a) we show the result for different initial weights $w^{(2)}_1$, in (b) we have rescaled by these weights to show the approximately linear behavior. In (c) we show the difference to a linear scaling ansatz for the largest initial weights and in (d) we show that this difference is to good approximation quadratic in the initial weight. These figures are taken from ref.\ \cite{DelZanna:2013eua}.}
\label{fig1}
\end{center}
\end{figure}

One can draw from these considerations an interesting conclusions for the comparison of large and small systems resulting from nucleus-nucleus (PbPb) and proton-nucleus (pPb) collisions. In order for $v_m\{n\}$ to be equal in the two cases, as is favored by the experimental results \cite{Chatrchyan:2013nka} (see also\cite{Basar:2013hea}), one needs for linear dynamics parametrically
\begin{equation}
\frac{S_{(m)l}{|}_\text{pPb}}{S_{(m)l}{|}_\text{PbPb}} = \left( \frac{N_\text{pPb}}{N_\text{PbPb}} \right)^{1-\frac{1}{n}}.
\end{equation}
Since the left hand side is independent of $n$ this can only be true for several $n$ if the number of sources agrees in the two cases, $N_\text{pPb} = N_\text{PbPb}$. In that case one has to conclude that the linear dynamic response functions are of equal size for pPb and PbPb which might come as a surprise given the differences in the collision geometry and center-of-mass energy. We plan to use our formalism for a more detailed investigation of this point in the future.

So far we have simply assumed that a perturbative expansion in deviations from a smooth and symmetric background is possible. Although this is a plausible assumption given the relative short life time and small spatial extent of the fluid produced by heavy ion collisions, it is not trivial. There might in principle be instabilities or turbulent behavior or the non-linear corrections could be too large for the scheme to be useful in praxis. In the following we describe results from the full numerical solution of the fluid dynamic evolution equations that test the validity and convergence of a perturbative expansion. 

We have initialized the hydro solver ECHO-QGP \cite{DelZanna:2013eua} with an enthalpy density as in eq.\ \eqref{eq1} with one or several coefficients $w^{(m)}_l$ non-zero. From the numerical solution we have then extracted the $m$'th azimuthal Fourier coefficient as a function of time $\tau$ and radius $r$,
\begin{equation}
\tilde w^{(m)}(\tau, r) = \frac{1}{w_\text{BG}(r)}\frac{1}{2\pi}  \int d\phi\;  e^{-im\phi} \, w(\tau,r,\phi).
\label{eq7}
\end{equation}
Fig. \ref{fig1} (a) shows the result for $\tilde w^{(2)}(\tau,r)$ at time $\tau=\tau_0 + 10 \, \text{fm/c}$ as a function of radius. The different curves show the result obtained for different values of the initial weight $w^{(2)}_1$. The range of these values is typical for a realistic Monte-Carlo Glauber model although most events have actually weights around $|w^{(2)}_1| \approx 0.2$ \cite{Floerchinger:2013vua}. 
One can now compare this numerical result to the expectation from perturbation theory. The lowest, linear order predicts that $\tilde w^{(2)}(\tau,r)$ scales linearly with the weight $w^{(2)}_1$. In fig.\ \ref{fig1} (b) we plot the rescaled curves and the fact that they are almost on top of each other shows that linear perturbation theory is already a very good approximation. Moreover, the small deviations from the linear scaling for larger weights $w^{(2)}_1$ that we plot again in fig.\ \ref{fig1} (c), can be understood as originating from the next order in perturbation theory where a contribution to $\tilde w^{(2)}(\tau, r)$ is allowed by symmetry reasons, namely cubic order. This is shown by the fact that the curves of fig.\ \ref{fig1} (c) fall again on top of each other when rescaled by $(w^{(2)}_1)^2$ as done in fig.\ \ref{fig1} (d). We have made similar tests also for more than one initial weights being non-zero and found in all situations that the fluid dynamic behavior is close to linear in the sense that the deviations from a linear evolution are small enough to be understood perturbatively as characteristic ``overtones'' \cite{DelZanna:2013eua}.

In summary, we have suggested a systematic perturbative expansion in deviations from a smooth and symmetric background solution for the fluid dynamics of heavy ion collisions. This approach leads to interesting insights into the structure of the fluid dynamic response to initial density perturbations and the perturbative series seems to have good convergence properties. The formalism can be used in praxis to connect initial fluctuations and experimentally accessible harmonic flow coefficients \cite{Floerchinger:2013rya} and we expect that it will provide useful for a quantitative determination of the thermodynamic and transport properties of the quark-gluon plasma.







\end{document}